\documentclass[aps,prl,twocolumn,a4paper,floatfix,showpacs,superscriptaddress]{revtex4}
\usepackage{graphicx}
\usepackage{hyperref}
\usepackage{color}
\usepackage{amsmath}
\usepackage[normalem]{ulem}
\usepackage{lineno}

\newcommand{\CRS}{CeRh$_2$Si$_2$ }
\newcommand{\ket}[1]{\left| #1\right\rangle}   
\begin{document}

\title{First-principles study on the role of surface in the heavy fermion compound CeRh$_2$Si$_2$}

\author{Yue-Chao Wang}
\affiliation{Laboratory of Computational Physics, Institute of Applied Physics and Computational Mathematics, Beijing 100088, China}

\author{Yuan-Ji Xu}
\affiliation{Beijing National Laboratory for Condensed Matter Physics, Institute of Physics, Chinese Academy of Science, Beijing 100190, China}

\author{Yu Liu}
\email{liu\_yu@iapcm.ac.cn}
\affiliation{Laboratory of Computational Physics, Institute of Applied Physics and Computational Mathematics, Beijing 100088, China}

\author{Xing-Jie Han}
\affiliation{Beijing National Laboratory for Condensed Matter Physics, Institute of Physics, Chinese Academy of Science, Beijing 100190, China}

\author{Xie-Gang Zhu}
\affiliation{Science and Technology on Surface Physics and Chemistry Laboratory, Jiangyou, Sichuan 621908, China}

\author{Yi-feng Yang}
\affiliation{Beijing National Laboratory for Condensed Matter Physics, Institute of Physics, Chinese Academy of Science, Beijing 100190, China}
\affiliation{School of Physical Sciences, University of Chinese Academy of Sciences, Beijing 100190, China}
\affiliation{Songshan Lake Materials Laboratory, Dongguan, Guangdong 523808, China}

\author{Yan Bi}
\affiliation{Center for High Pressure Science and Technology Advanced Research, Beijing 100094, China}

\author{Hai-Feng Liu}
\affiliation{Laboratory of Computational Physics, Institute of Applied Physics and Computational Mathematics, Beijing 100088, China}

\author{Hai-Feng Song}
\email{song\_haifeng@iapcm.ac.cn}
\affiliation{Laboratory of Computational Physics, Institute of Applied Physics and Computational Mathematics, Beijing 100088, China}

\date{\today}

\begin{abstract}
In the heavy fermion materials, the characteristic energy scales of many exotic strongly correlated phenomena (Kondo effect, magnetic order, superconductivity, etc.) are at milli-electron-volt order, implying that the heavy fermion materials are surface sensitive. Here, we investigate the electronic structures for Si- and Ce-terminated surfaces of CeRh$_2$Si$_2$ by first-principles methods. Our research reveals three notable impacts of surface effects on electronic structures, which are consistent with recent angle-resolved photoemission spectroscopy (ARPES) experiments. Firstly, the relaxation of surface crystal structures changes the relative position of Fermi level, adjusts the dispersion of bands and enhances the Kondo resonance. Secondly, the decrease of the hybridization between the Ce-4$f$ and conduction electrons in the surface layer leads to a weaker Kondo resonance peak and the shift of spin-orbit bands. Thirdly, the variation of crystal electric field around surface Ce atoms affects the splitting of Kondo resonance peaks, and also pushes down the lower-Hubbard bands of surface 4$f$ electrons. Moreover, we find the characteristic of bulk's lower-Hubbard bands, which was overlooked in previous works. Our investigation suggests that these surface effects are potentially important and highlighted in the future researches on properties of strongly correlated materials.

\end{abstract}

\maketitle

\section{Introduction}
Cerium-based compounds have many exotic and interesting properties, such as heavy fermion behavior, superconductivity, magnetic order, which are believed to be originated from the strongly correlated 4$f$ electrons and its hybridization with the conduction electrons \cite{Sigrist1991,Stewart1984,Budko1996,Misra2007,Hewson1993,Movshovich1996,Gegenwart2008}.
Among these compounds, CeRh$_2$Si$_2$ has been extensively studied for its strong crystal electric field and anisotropic crystal structure \cite{Kawarazaki2000,Graf1998,Boursier2008,Patil2016,Vildosola2005,Lu2018,Radwanski2019,Pourret2017}. The de Haas-van Alphen and neutron scattering techniques have been used to reveal the hybridization of $f$-electrons with conduction electrons ($c$-electrons) \cite{Kawarazaki2000,Severing1989,Araki2001}. \textcolor{black}{More importantly, the layered structure of CeRh$_2$Si$_2$ single crystal can be technically cleaved with different terminated atoms, and it makes CeRh$_2$Si$_2$ one of the excellent candidates for investigation of surface properties\cite{Vyalikh2010,Danzenbacher2011,Hoppner2013,Chikina2014}. With the recent progress of ARPES technique which is believed as a surface sensitive approach\cite{Lv2019,Zhang2018,Zhu2020,Feng2018}, the high-quality experimental results of electronic structures from Si- and Ce-terminated CeRh$_2$Si$_2$ surfaces were reported \cite{Patil2016,Poelchen2020}.}
In these experimental results, the strength of Kondo resonance peak, Kondo temperature and other fine structure around Fermi level show notable differences between the samples with Si- and Ce-terminated surfaces, which indicates the hybridization strength and crystal electric field are affected by the different surface environments. These experimental phenomena have shown that the environment of surfaces can be of great difference and the surface have significant impact on the electronic structures. \textcolor{black}{However, no comprehensive interpretation has been given for the physical mechanism about how the surface affects electronic structures of CeRh$_2$Si$_2$ from experimental side. }

On the theoretical side, many studied focused on the bulk properties and have been done by model Hamiltonian approaches and first-principles simulations \cite{Radwanski2019,Vildosola2004,Vildosola2005,Lu2018}.
The equilibrium volume, $c/a$ ratio and bulk modulus of CeRh$_2$Si$_2$ are obtained from density functional theory (DFT) calculations \cite{Vildosola2004}. The crystal electric field effect and Kondo resonance of bulk 4$f$-electrons are studied by density functional theroy plus dynamical mean-field theory (DFT+DMFT) \cite{Vildosola2005,Lu2018}. All these theoretical works have discribed some phenomena of CeRh$_2$Si$_2$ properly, such as the anisotropic hybridization interaction, splitting of Kondo peak and mixed valence Ce atomic configuration. However, most of these works concentrate on the bulk of CeRh$_2$Si$_2$, and the theoretical investigations focusing on its surface influence are rare. \textcolor{black}{A recent work simulated the surface electronic structures of CeRh$_2$Si$_2$ with DFT\cite{Poelchen2020}, and the results show the proper $spd$ dispersion compared with recent ARPES experiments\cite{Patil2016,Poelchen2020}. However, the important strongly correlated electron features are lacking in DFT, which makes it unable to investigate the influence of surface on these features that have been observed in ARPES experiments \cite{Patil2016,Poelchen2020}. Therefore, further theoretical investigations with a proper treatment of strongly correlated effect to reveal the role of surface in CeRh$_2$Si$_2$ is necessary.}

\textcolor{black}{In this work, we perform a investigation on the surface effect of CeRh$_2$Si$_2$ with first-principles simulations, in which the strongly correlated effect and surface structure is explicitly considered by DMFT approach and slab models. Three aspects of the electronic properties of CeRh$_2$Si$_2$ with different terminated surfaces are studied systematically. }
Firstly, we focus on the surface crystal structure relaxation. The DFT+DMFT simulations of band structures are performed on relaxed surface crystal structures of Si- and Ce-terminated surfaces.
Compared to the unrelaxed crystal structures, the relaxed crystal structures give band structures (relative position of Fermi level and bands' shape) similar to APRES. We show that this is caused by the different distribution of the electron density due to relaxation.
Secondly, we investigate the effect of hybridization on the surface, and we reveal the difference between surface and bulk electronic properties in Si- and Ce-terminated surfaces respectively for the decreasing of $c$-$f$ hybridization strength on the surface. Thirdly, we take the crystal electric field effect into consideration and reproduce the electronic structures from ARPES results successfully, and find the surface electronic information can cover the main feature from experiment, which implies surface states matter in the first-principles simulations of CeRh$_2$Si$_2$.

The paper is organized as follows. In Sec.II, we introduce the methods and parameters used in this paper. In Sec.III, the results of Si- and Ce-terminated slab model are exhibited, and the analysis on the surface effect on the electronic properties of CeRh$_2$Si$_2$ is performed. Sec.IV closes the the paper by a summary of the main findings of this work and some general remarks.

\section{Method}

\begin{figure}
\centering
\includegraphics[width=0.48\textwidth]{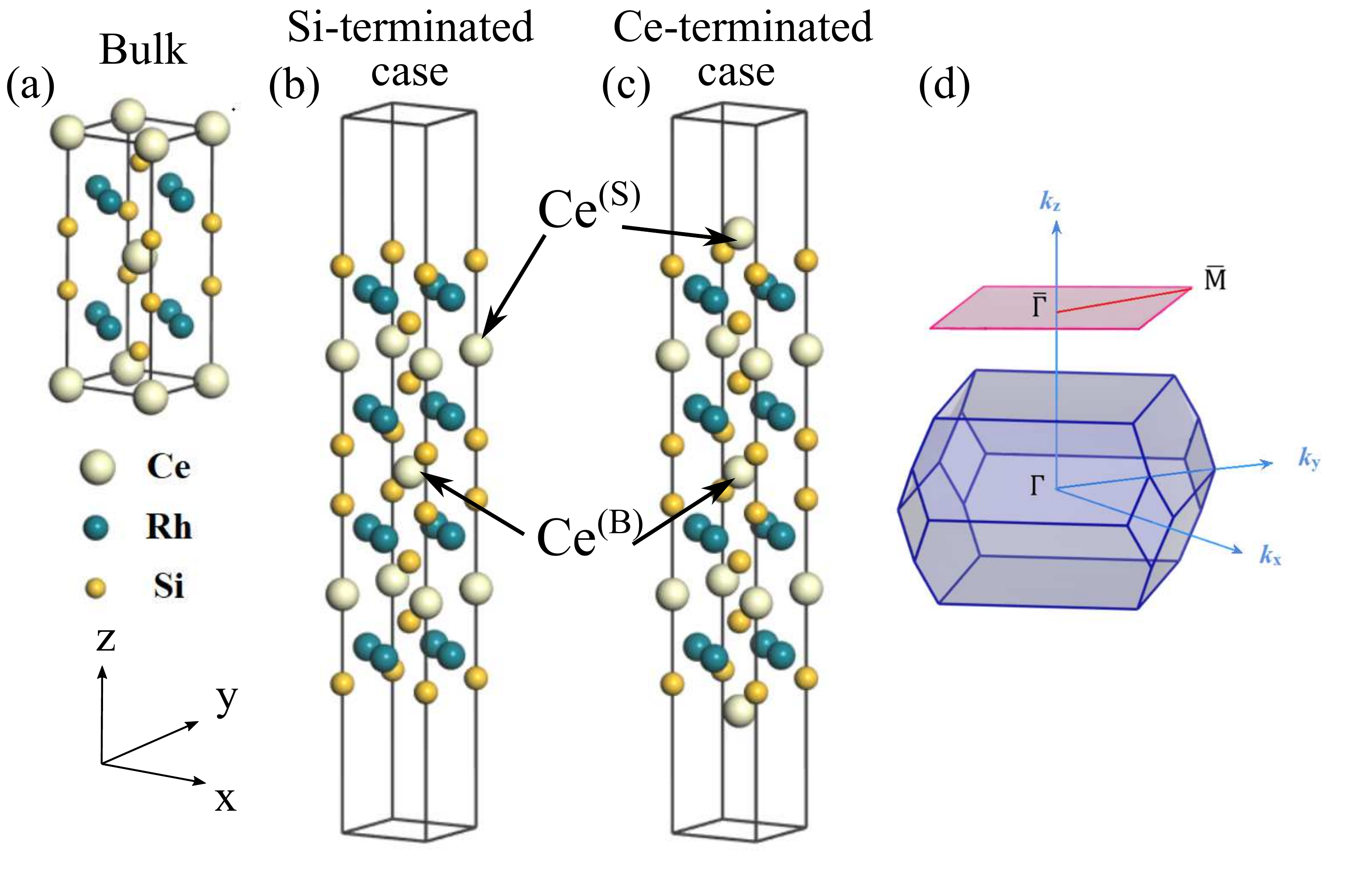}
\caption{\label{fig:str}The crystal structure of CeRh$_2$Si$_2$. The left panel shows the cases of  bulk (a), Si-terminated slab (b) and Ce-terminated slab (c). The exposed surface in both slab models is (100) direction. We point out the surface Ce and bulk Ce referred in our work with label Ce$^{(\text{S})}$ and Ce$^{(\text{B})}$ in both slabs. The right panel (d) is the first Brillouin zone of the bulk (blue) and surface (red). }
\end{figure}

For crystal structure relaxation, the DFT+$U$ method is exploited for both Si- and Ce-terminated cases as a relaxation of slabs on DFT+DMFT is expensive\cite{Himmetoglu2014,Anisimov2010book,Yang2018}, and the validity of this treatment is discussed in our Supplemental Material\cite{Suppl}. All the exchange-correlation functional used in the this work is the conventional Perdew-Burke-Ernzerhof (PBE) functional \cite{Perdew96PBE}. We consider the spin-orbit coupling during our simulation, while long-range magnetic orders are not considered in the simulations. The DFT+$U$ simulation is performed in the Vienna Ab initio Simulation Package (VASP) code with the projector augmented wave (PAW) method\cite{Kresse1996}. The $f$ electrons is treated as valence electrons in PAW pseudopotential, and a plane wave energy cut of 350 eV is used. For the simulation of slabs, the k-mesh is set to $25\times25\times1$, and the Gaussian smearing is used to avoid sample error along $k_{z}$ direction. We have also tested the density of states (DOS) results on a $31\times31\times1$ k-mesh, and no obvious difference is found from $25\times25\times1$ ones. The on-site interaction parameters, Hubbard $U$ and Hunds exchange $J$, used for DFT+$U$ are $U$ = 6.0 eV and $J$ = 0.7 eV which is a conventional choice\cite{Vildosola2004,Vildosola2005,Lu2018}. As shown in Fig.\ref{fig:str}, \CRS crystal takes a body-centered tetragonal ThCr$_2$Si$_2$-type structure belonging to the D$_{4h}$ point group (space group I4/mmm No. 139). The lattice parameters of \CRS bulk is fixed as experimental results\cite{Grier1984}. The slab models used in this work for Si- and Ce-terminated (100) surfaces are also displayed in Fig.\ref{fig:str}. The vacuum added on the slab is 15 angstrom, and the convergency is tested by the DOS calculation with a 20 angstrom vacuum. During the relaxation, the central layer Ce, and the Si-Rh layers beside it are fixed at the bulk position to simulate the bulk, and the lattice parameters inside the surface is kept as the experiment ones\cite{Grier1984}. The first Brillouin zone of the bulk and surface and the high symmetric points used in this work are shown in Fig.\ref{fig:str}(d).
As shown in Fig.\ref{fig:str}(a-c), the Ce atoms in the surface and bulk are marked by Ce$^{(\text{B})}$ and Ce$^{(\text{S})}$ respectively.
If not specified, all the simulations in this work are performed with relaxed structures.

The electronic structures simulations in isotropic environment (including band structures and density of states) are done with a charge fully self-consistent DFT+DMFT calculation \cite{Kotliar2006,Georges1996,Kotliar2004}. The DMFT part is solved by the eDMFT software package, and DFT part is performed in WIEN2k software package \cite{Haule2010,Wien2k}. The DFT performed in WIEN2k is based on the the full potential linearized augmented plane-wave method (LAPW), with $R_\text{mt}K_\text{max}$ = 8.0 and muffin-tin radii, $R_\text{mt}$, 2.5 a.u. for Ce, 2.4 a.u. for Rh and 2.0 a.u. for Si. The multi-orbital Anderson impurity model is solved by the hybridization expansion continuous-time quantum Monte Carlo impurity solver (CTQMC)\cite{Werner06,Haule2007}. The temperature is $T =$ 50 K and the Hilbert space of atomic eigenstates is truncated into electron occupancy from 0 to 3. The crystal structures, k-points and on-site $U$ and $J$ values are the same as relaxation part.

The electronic structure simulations with crystal electric field effect are perform based on the isotropic simulation.
The crystal parameters are calculated from a constrained-DFT approach introduced by Nov${\rm\acute{a}}$k with LAPW method in WIEN2k, and it is used to avoid the self-interaction error in DFT \cite{Novak2013,Mihokova2015}. The electron of $4f$ is constrained to one on Ce atoms, and the radial part local orbital of 4$f$ is the same as in DFT+DMFT. Other parameters are kept the same as those mentioned above.

\section{Results and Discussion}

\subsection{Surface structure relaxation}
The first issue that we want to address is the importance of including the surface crystal structure relaxation in the electronic structures calculations\cite{Mandal2017}. The concept of surface crystal structure relaxation refers to the change of geometric structure of surface caused by the unbalanced force performed on the surface layer atoms. After relaxation, in the Si-terminated case, the surface Si layer moves into
the center by 0.204 \AA, and the outmost Ce layer moves into the center by 0.064 \AA. In the Ce-terminated case, the surface Ce layer moves toward the center by 0.186 \AA. Consequently, the external potentials are changed by different surface atomic structures and we will see below that the electronic band structures are found to be significantly influenced before and after relaxation.

\begin{figure}
\centering
\includegraphics[width=0.48\textwidth]{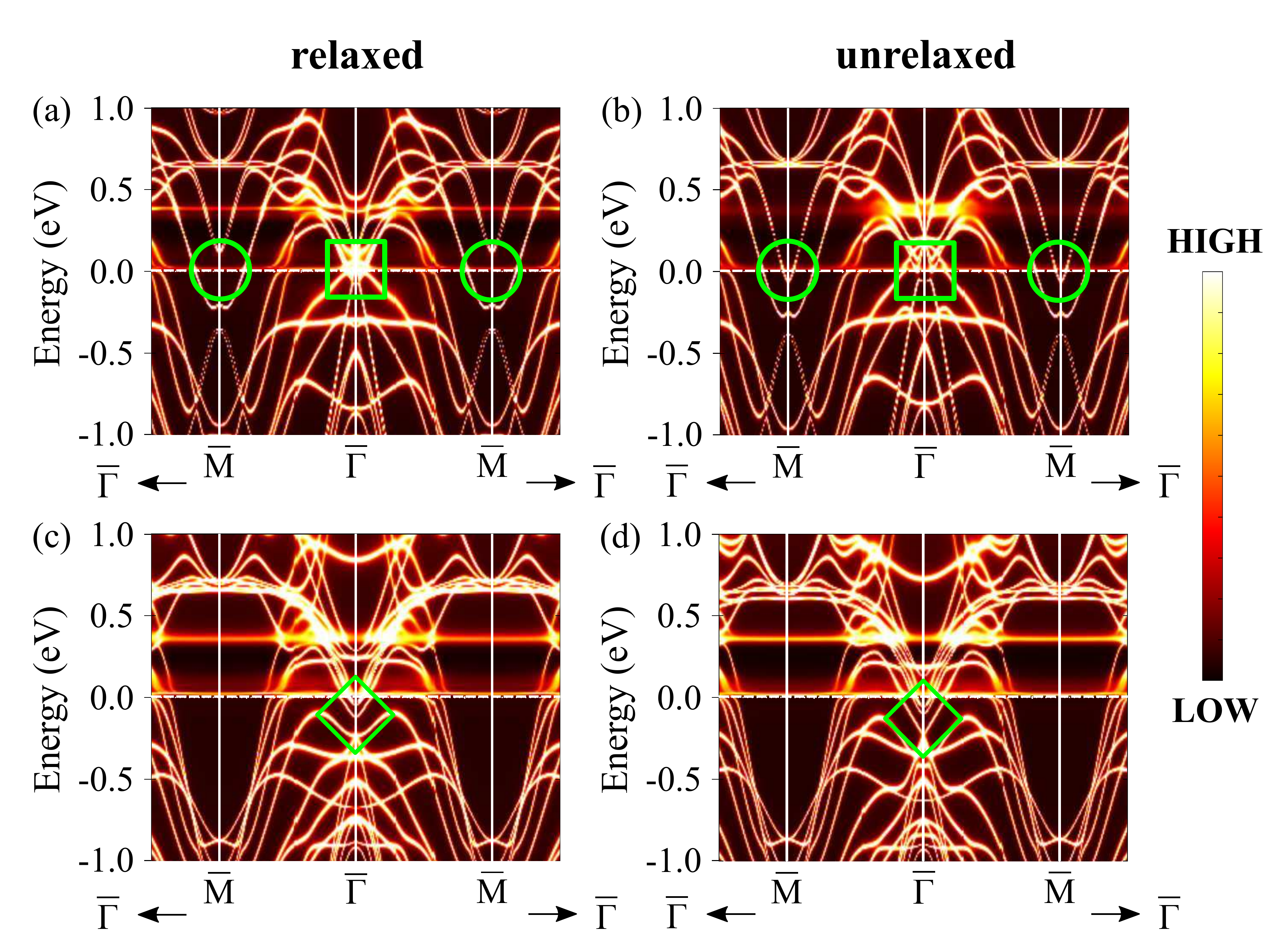}
\caption{\label{fig:arpes}The momentum-resolved spectral functions calculated by DFT+DMFT.
 (a) and (c) are the results of relaxed crystal structures of Si- and Ce-terminated cases respectively. (b) and (d) are the results of unrelaxed crystal structures of Si- and Ce-terminated cases respectively.
 The green frames point out the regions with main differences between the band structures with and without the relaxation.
 \textcolor{black}{Zoom-in figures around $\bar{\Gamma}$ point of (a) and (c) are also shown in Supplementary Materials Fig.S4\cite{Suppl}.}}
\end{figure}

Fig.\ref{fig:arpes} compares the calculated electronic band structures for the relaxed and unrelaxed crystal structures along the high-symmetry line of the Brillouin zone. Quantitative differences can be observed for the low-energy spectral functions around the Fermi level. For the relaxed Si-terminated surface in Fig.\ref{fig:arpes}(a), we observe a hole-like, linearly dispersing surface resonant band (square region) with a cusp around the Fermi level at $\bar{\Gamma}$ point. This cusp is shifted much above the Fermi level for the unrelaxed case as shown in Fig.\ref{fig:arpes}(b). On the contrary, the location of the electron-like band (the circle region) around the $\bar{\text{M}}$ point for the relaxed structure moves above the Fermi level. In both cases, the experimentally observed Shockley-type surface states around -0.5 eV at $\bar{\text{M}}$ point can be well reproduced\cite{Patil2016}, which may originate from the dangling bonds of surface Si. For Ce-terminated surface as shown in Fig.\ref{fig:arpes}(c,d), the most apparent discrepancy appears around -0.25 eV at $\bar{\Gamma}$ point (diamond region). The crossing point observed in the unrelaxed case is separated in the relaxed case. The reported rocket-shaped features below -0.5 eV are also shown in our calculations for both cases\cite{Patil2016}. Furthermore, we find that our results after relaxation are consistent with experimental results\cite{Patil2016}. This implies that relaxation plays an important role in calculating the electronic band structures of CeRh$_2$Si$_2$.

Besides the band dispersion around Fermi level, it can also be observed that the $f_{7/2}$ band is enhanced with the surface relaxation. This feature \textcolor{black}{cannot} be obtained from ARPES results, as ARPES probes the information below Fermi level. Together with this feature, the Kondo peaks and spin-orbit bands are also enhanced sightly. The detailed results of DOS and hybridization strength are displayed in the Supplemental Materials Fig.S6\cite{Suppl}.

\begin{figure*}
\centering
\includegraphics[width=1.0\textwidth]{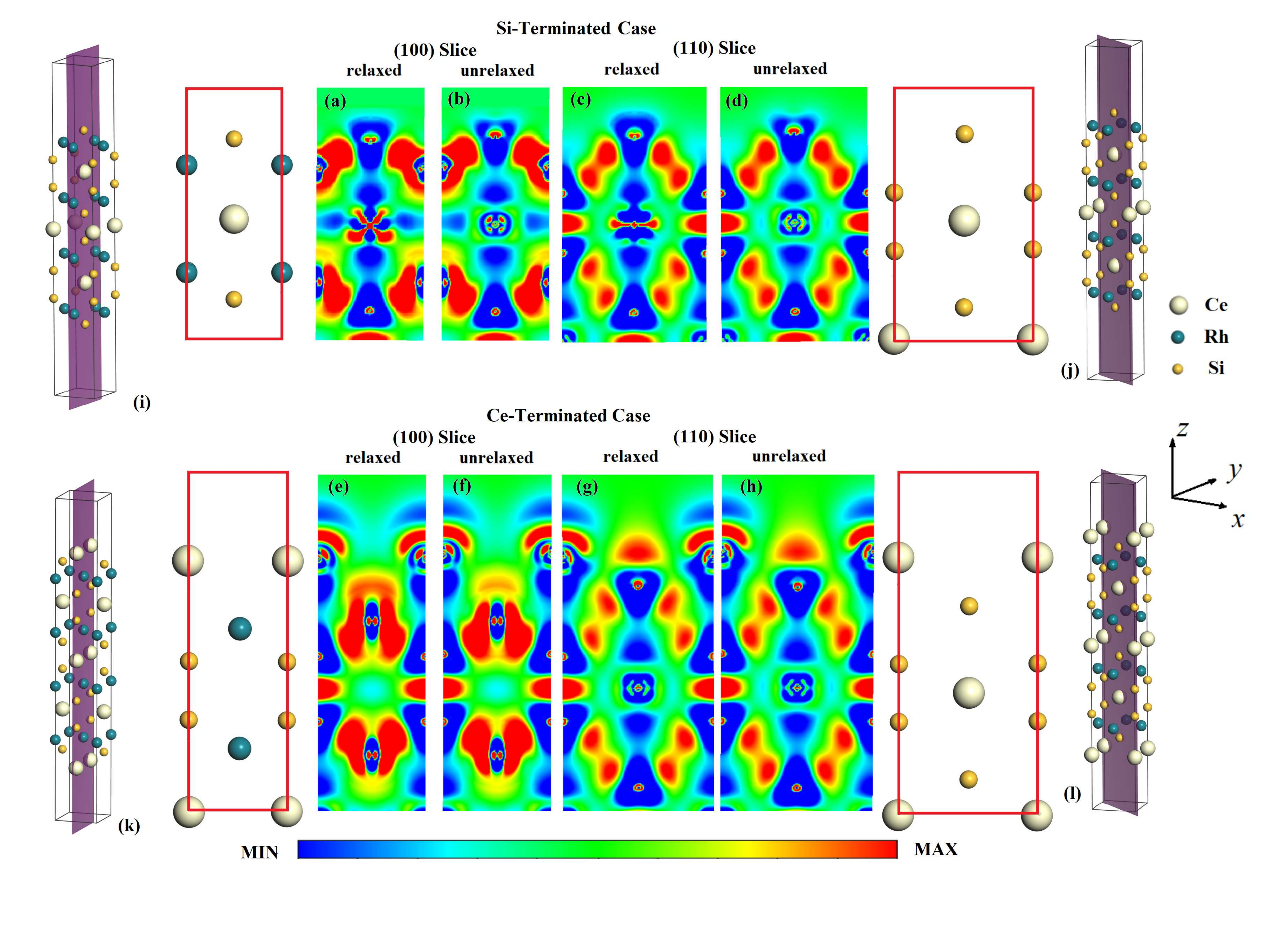}
\caption{\label{fig:doc}The difference of electron density between results of DFT+DMFT calculation and the superposition of spherical atomic electron density. The electron density differences of typical slice of Si- and Ce- terminated cases with and without relaxation are shown in the figure (a)-(h). The warm color means the increasing of electron density and cold color means the decreasing of electron density.
(i)-(l) show the definition of the slice used in (a)-(h). The position and direction of different slices are specified by violet layers, and atomic positions in (a)-(h) are shown in red frames.}

\end{figure*}

These differences can be qualitatively understood by the enhancement of bonding effect between the surface and sub-surface layer atoms. Fig.\ref{fig:doc} displays the electron density difference for Si-/Ce-terminated surfaces. In the Si-terminated case, there are more electrons in the range between Rh-Si layer and Ce layer after relaxation, making the bonding between Ce and Rh-Si layer stronger.
The enhancement of bonding effect will make the band with bonding characteristic move downwards and with anti-bonding characteristic move upwards in energy.
A similar but weaker effect can also be observed in Ce-terminated cases between the outmost Ce-layer and Rh-Si sub-layer. All these enhancements of bonding effect are directly related to the shorten of the distance between the surface and the bulk from relaxation.
The enhancement of $f_{7/2}$, Kondo peaks and spin-orbit bands may be partially attributed to the increasing of hybridization after the relaxation. However, it should be emphasized that the hybridization variation caused by surface itself is much notable than relaxation, which is investigated in the following part.

\textcolor{black}{A surface electronic band structure simulation work of CeRh$_2$Si$_2$ with DFT method is also performed by G. Poelchen \emph{et al}\cite{Poelchen2020}. The band structure of their work also showed a linear dispersion around $\bar{\Gamma}$ point in Si-terminated case which is different from Ce-terminated case as we have shown above. However, as their simulation is simulated on DFT level, features like Kondo peak, $f_{7/2}$ band and Hubbard band which are corresponding to the strongly correlated effect were not studied in their work. And as we have mentioned above, these features are important to understand the influence of surface effects.}

Here, it should be pointed out that the surface crystal structure relaxation is calculated at DFT+$U$ level, and the crystal structure is used as an approximation to the DFT+DMFT results. Because DFT and DFT+$U$ stand for the limit of weak and strong correlation respectively, we assume that the performance of crystal structure relaxation of DFT+DMFT will be between these two approaches. The validity of this approximation for CeRh$_2$Si$_2$ system is confirmed by the comparison of electronic structures results from DFT, DFT+$U$ and DFT+DMFT and the structure relaxation results of DFT and DFT+$U$. The detailed discussion of this part is in the Supplemental Material part II\cite{Suppl}.

\subsection{Hybridization strength}

In this section, we focus on the hybridization strength of the local $f$-electron and the conduction electrons in the surface area.
The hybridization strength between surface $f$- and $c$-electrons is anticipated to decrease due to the disappearance of $c$-electrons on one side of the surface. As a result, the electronic properties of Ce$^{(\text{S})}$ are different from Ce$^{(\text{B})}$ in CeRh$_2$Si$_2$. In Fig.\ref{fig:doc} the electron density between Si-Rh and Si-Si is shown to be much larger than other areas in the slab, indicating the strong bonding effect between Si-Rh and Si-Si.
The electrons of Si-Rh and Si-Si region make the skeleton of $c$-electron environment.
The electrons on Ce surrounded by this $c$-electron environment (like Ce$^{(\text{S})}$ in Si-terminated case and Ce$^{(\text{B})}$) decreases compared with isolate atom. It indicates a strong hybridization of electrons on Ce atom (like 4$f$ electrons) with $c$-electrons, which improves the energy of Ce atomic states and transfers the atomic electrons to $c$-electrons. On the contrary, there are more electrons concentrating on Ce$^{(\text{S})}$ in Ce-terminated case with a hemispherical distribution, which implies a weak hybridization and less variation from atomic state.
The relation between the hybridization strength and electron distribution in real space have also been observed with spectroscopic imaging scanning tunneling microscopy (SI-STM) in heavy fermion materials\cite{Hamidian2011}.

\begin{figure}
\centering
\includegraphics[width=0.48\textwidth]{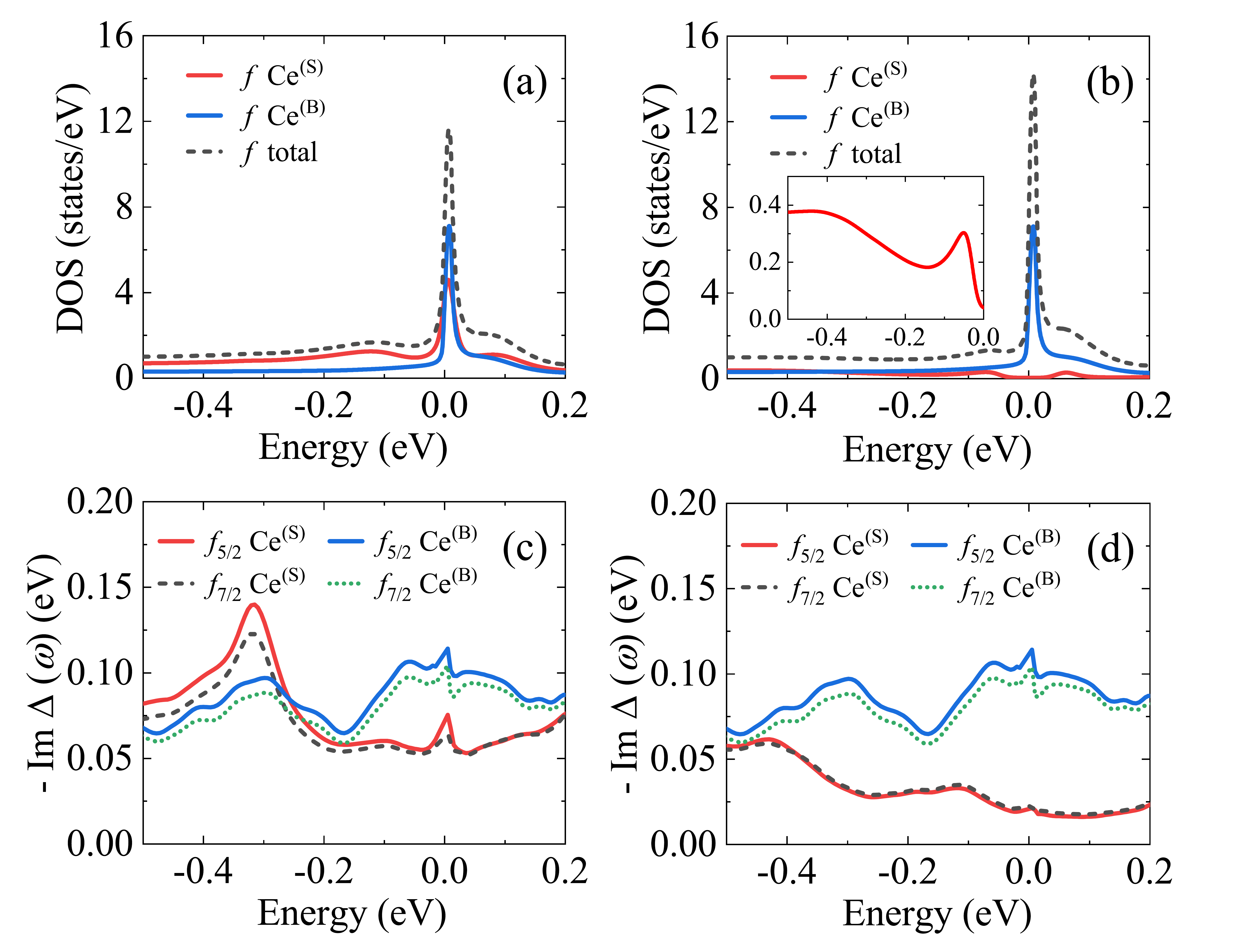}
\caption{\label{fig:hyb}Density of states and imaginary part of hybridization functions in Si- and Ce-terminated case. (a) and (b) are the DOS of Ce $4f$-electrons in Si- and Ce-terminated cases respectively. The red solid lines are the $4f$-electron DOS of Ce$^{(\text{S})}$. The blue solid lines are the $4f$-electron DOS of Ce$^{(\text{B})}$. The black dash lines are the total $4f$-electron DOS of Ce. The inset of (b) shows the $4f$-electron DOS of Ce$^{(\text{S})}$ in a smaller scale. (c) and (d) show the imaginary part of hybridization function of Si- and Ce- terminated case respectively on real frequency. The solid red and blue lines are for $f_{5/2}$ of Ce$^{(\text{S})}$ and Ce$^{(\text{B})}$. The black dash and green dot lines are for $f_{7/2}$ of Ce$^{(\text{S})}$ and Ce$^{(\text{B})}$.}
\end{figure}

The hybridization strength is closely related to the Kondo resonance peaks at Fermi level and a sharper Kondo peak generally corresponds to a stronger hybridization strength. To explicitly study the effect of hybridization in surface, we show the $4f$-projected DOS (PDOS) for Si- and Ce-terminated cases of CeRh$_2$Si$_2$ in Fig.\ref{fig:hyb}(a) and Fig.\ref{fig:hyb}(b). The bulk Kondo peaks are sharper than the surface Kondo peaks for both cases. In contrast to the single Kondo peak obtained in the Si-terminated case, the surface Kondo peak for the Ce-terminated case splits into two small Kondo peaks. The distribution of electron density around Ce atom is responsible for the distinct behaviors of the Kondo peaks. In Si-terminated case, the Ce$^{(\text{S})}$ still has an environment similar to the bulk. However, for Ce-terminated case, the Ce$^{(\text{S})}$directly exposes to the vacuum which makes it closer to an isolated atom, thus, the Kondo peak vanishes.

The hybridization strength is directly related to the imaginary part of the hybridization function\cite{Jacob2010,Haule2014,Huang2019}. The definition of hybridization function is
\begin{equation}
\begin{aligned}
\label{eq:hyb}
&\Delta(\omega) = \sum_{k}\frac{|V_k|^2}{\omega-\epsilon_k+i\eta},\\
\end{aligned}
\end{equation}
where $V_k$ and $\epsilon_k$ represent the hybridization parameter and the dispersion of the conduction electrons, respectively. $\eta\rightarrow0^+$ is an infinitesimal positive real number. The imaginary part of hybridization function, which can be used to characterize the hybridization strength, is obtained from the imaginary part of the hybridization function
 \begin{equation}
\begin{aligned}
\label{eq:hyb1}
\text{Im} \Delta(\omega) = -\pi\sum_{k}|V_k|^2\delta(\omega-\epsilon_k).\\
\end{aligned}
\end{equation}
Fig.\ref{fig:hyb}(c) displays $\text{Im} \Delta(\omega)$ for the $4f$ Ce$^{(\text{S})}$ atom and the Ce$^{(\text{B})}$ atom in Si-terminated case. The absolute values of the peak at $\omega=0$ for the Ce$^{(\text{B})}$ atom is 38\% larger compared to the $4f$ Ce$^{(\text{S})}$ atom. For the Ce-terminated case shown in Fig.\ref{fig:hyb}(d), $\text{Im} \Delta(\omega)$  for the $4f$ Ce$^{(\text{S})}$ around the Fermi level is very small and the absolute value is only 21\% compared with Ce in bulk. The results from $\text{Im} \Delta(\omega)$  are consistent with PDOS, and our results suggest that the electronic structures at Fermi level and the hybridization strength are closely related.


\textcolor{black}{According to previous researches \cite{Reinert2001,Patthey1985,Vildosola2005}, the position of spin-orbit side peak is correlated to the hybridization strength of $f$-$c$ electrons. Stronger hybridization could make the spin-orbit side peak more close to the Fermi level. As shown in Fig.\ref{fig:hyb}(a) and (b), the spin-orbit side peak of Si-terminated case is more close to the Fermi level than that of Ce-terminated case which is in qualitative agreement with ARPES experiments\cite{Patil2016,Poelchen2020}. Thus the difference between position of spin-orbit side peaks can also imply a stronger hybridization in Si-terminated case than that in Ce-terminated case. However, some other factors, such as total electron number and crystal electric field, can also contribute to this difference. Thus, the position of spin-orbit side peak in this work is just a qualitative evidence for the strength of hybridization.}

\subsection{Crystal electric field effect}

\begin{figure}
\centering
\includegraphics[width=0.48\textwidth]{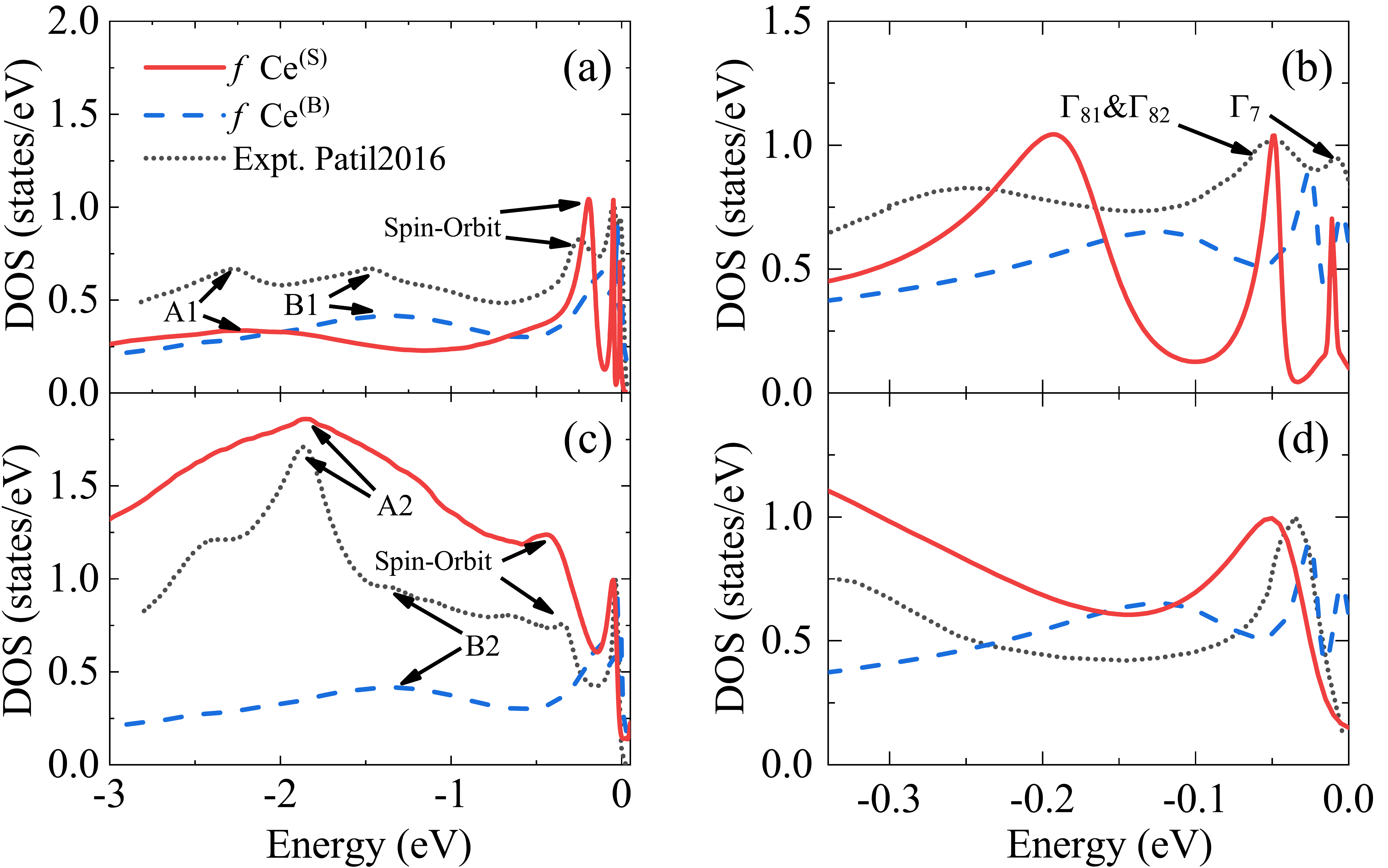}
\caption{\label{fig:dos}The DOS from DFT+DMFT simulation with crystal electric field effect.
(a) and (c) show the DOS of Si-terminated and Ce-terminated cases respectively. (b) and (d) are their enlarged views around Fermi level. The experimental results from Patil are also shown for comparison \cite{Patil2016}. The peaks of spin-orbit coupling, $\Gamma_7$, $\Gamma_{81}$ and $\Gamma_{82}$ are pointed out directly in the figure. A1 and B1 stand for the peaks of lower Hubbard band of surface and bulk in Si-terminated case respectively, and so do A2 and B2 for Ce-terminated case. All the DOS data are renormalized respected to the highest resonance state at around Fermi level (at about -0.01 eV) from experimental results respectively.
}
\end{figure}

In the presence of crystal electric field, the $f_{5/2}$ orbitals split into $\Gamma_7$, $\Gamma_{81}$ and $\Gamma_{82}$ oritals. In this section, $\Gamma_7$, $\Gamma_{81}$ and $\Gamma_{82}$ are defined as
\begin{equation}
\begin{aligned}
\label{eq:gamma7}
\Gamma_{7}=a\ket{\frac{5}{2};\pm\frac{5}{2}}&-b\ket{\frac{5}{2};\pm\frac{3}{2}},\\
\Gamma_{82}=b\ket{\frac{5}{2};\pm\frac{5}{2}}&+a\ket{\frac{5}{2};\pm\frac{3}{2}},\\
\Gamma_{81}=\ket{\frac{5}{2};\pm\frac{1}{2}}.\\
\end{aligned}
\end{equation}
Here the parameters $a=\sqrt{1/6}$ and $b=\sqrt{5/6}$ are chosen as is proposed in other theoretical works\cite{Lea1962,Vildosola2005}. Fig.\ref{fig:dos} shows the $f$-bulk and $f$-surface DOS for Si-terminated and Ce-terminated cases with crystal electric field. The experimental results from the integrated resonance enhanced ARPES are also given for comparison. The peaks referring to $\Gamma_7$, $\Gamma_{81}$ and $\Gamma_{82}$ are marked on the figure\cite{Patil2016}.

\begin{table}[b]
\caption{\label{tab:cry}Crystal electric field (CEF) parameters from constrained-DFT calculation. The absolute value of $\Gamma_7$ is given in the second column. The values of $\Gamma_{81}$ and $\Gamma_{82}$ referring to $\Gamma_7$ are given in the third and the forth column. The definitions of $\Gamma_7$, $\Gamma_{81}$ and $\Gamma_{82}$ are the same as Eq.(\ref{eq:gamma7}).
For comparison, the experimental and previous theoretical values are also given\cite{Patil2016,Vildosola2005}.}
\begin{tabular*}{0.45\textwidth}{@{\extracolsep{\fill}} l|lll}
\hline\hline
CEF parameters (eV)  &$\Gamma_7$ & $\Gamma_{81}-\Gamma_{7}$ & $\Gamma_{82}-\Gamma_7$\\
 \hline
Ce$^{(\text{S})}$ in Si-Case           &0.496  &0.055  &0.081  \\
Ce$^{(\text{S})}$ in Ce-Case           &0.063  &0.004  &0.009  \\
Ce$^{(\text{B})}$ in Bulk                  &0.691  &0.049  &0.084  \\
Expt. in Si-Case\cite{Patil2016}           &--     &0.048  &0.062  \\
DFT  in  Ce-bulk\cite{Vildosola2005} &--   &0.021   &0.048 \\
\hline\hline
\end{tabular*}

\end{table}

$\Gamma_{81}$ and $\Gamma_{82}$ states are merged in our bulk and surface results shown in Fig.\ref{fig:dos}(b) in Si-terminated case. The similar mergence is reported in previous theoretical and experimental works when temperature is higher than 30 K\cite{Vildosola2005,Patil2016}. For the surface $f$ electrons, the splitting between the peak of $\Gamma_7$ and the merged peak, which is composed of $\Gamma_{81}$ and $\Gamma_{82}$ peaks, within -0.1$\sim$0 eV is around 60 meV. And the peak caused by spin-orbit coupling is around -0.25 eV. All these low-energy features are in good agreement with previous studies\cite{Poelchen2020,Patil2016}.
In contrast, for the bulk $f$-electrons, the resonance peaks ($\Gamma_7$, $\Gamma_{81}$ and $\Gamma_{82}$) have a smaller splitting energy and the deviation of the spin-orbit peak from the experimental results is more significant.

For the high-energy part, contrary to the experimental results that there are two peaks located around -2.0 eV (Fig.\ref{fig:dos}(a) mark A1) and -1.5 eV (Fig.\ref{fig:dos}(a) mark B1), only one peak is generated in our calculations for the surface and bulk cases. The positions of the high-energy peaks for bulk and surface are close to the locations of the two peaks observed experimentally, indicating that the APRES results are composed of both surface and bulk properties of the material at high-energy.

Fig.\ref{fig:dos}(c) and (d) show the results of the Ce-terminated case. For the Ce$^{(\text{S})}$, the splitting of the Kondo peak is not observed due to the fact that $\Gamma_7$, $\Gamma_{81}$ and $\Gamma_{82}$ orbitals are near degenerate. It can be understood from the crystal
electric field parameters obtained from constrained-DFT calculations listed in Table.\ref{tab:cry}. The energy differences among $\Gamma_7$, $\Gamma_{81}$ and $\Gamma_{82}$ are quite small. However, one can readily see from Table.\ref{tab:cry} that this near-degeneracy is lifted for Ce$^{(\text{B})}$. Then, the peak is splitted again in this case. The spin-orbit peak for Ce$^{(\text{S})}$ is closer to the experimental results within -0.3$\sim$-0.4 eV. In Fig.\ref{fig:dos}(c), the experimentally observed Hubbard peak at position A2 around -2.0 eV is also obtained in our calculation for Ce$^{(\text{S})}$. The broadening of the calculated Hubbard peak can be attributed to the fact that the calculated DOS contains information of all k-spacing while the experimental DOS is integrated along a certain k-path. For position B2 around -1.3 eV, the experimentally observed hump can be attributed to the contributions from Ce$^{(\text{B})}$.

The distinction of crystal electric field in different surfaces can be quantitatively accounted for by the crystal electric field parameters listed in Table.\ref{tab:cry}. The absolute values and splitting energies between the triplet splitting $f_{5/2}$ states in bulk and different surfaces are given.
It is clear that the absolute values of Ce$^{(\text{S})}$ atoms for the Ce-terminated case is one order less than that of bulk. As the absolute value are referring to the Fermi energy of corresponding slabs, the smaller value of Ce$^{(\text{S})}$ implies a weaker crystal electric field compared with Ce$^{(\text{B})}$.
It is reasonable that the Ce$^{(\text{S})}$ in Ce-terminated case experience the weakest crystal electric field effect in the absence of half Si-Rh crystal environment. The \textcolor{black}{difference} in the absolute values of Ce$^{(\text{S})}$ and Ce$^{(\text{B})}$ also have impacts on the 4$f$ electron levels, and contribute to the shift of Hubbard bands of Ce$^{(\text{S})}$ comparing to the Ce$^{(\text{B})}$. Meanwhile, the splitting between states in Ce-terminated case is also one order less than the other cases which implies that the behavior of the Ce$^{(\text{S})}$ in Ce-terminated case is similar to the isolate atom. This could also explain the near degeneracy of $f_{5/2}$ in Ce-terminated case. A slight enhancement (about 6 meV) of CEF splitting on Ce$^{(\text{S})}$ of Si-terminated case comparing with Ce$^{(\text{B})}$ is also be observed in constrained-DFT results. It may have contribution to the larger splitting between peaks of $\Gamma_7$ and $\Gamma_{81}$\&$\Gamma_{82}$ of Ce$^{(\text{S})}$ in Si-terminated case. We suppose this variation of CEF splitting may originate from the asymmetric charge distribution of Ce$^{(\text{S})}$ in Si-terminated case.

We have shown that the main features from resonance enhanced ARPES results can be well reproduced by our simulation of surface 4$f$ electrons. Despite the much lower height of resonance peak comparing to the bulk around Fermi level of surface $f$ state as shown in the Fig.\ref{fig:hyb}, most of the low-energy information obtained by ARPES can be traced back to the Ce$^{(\text{S})}$ electron states, especially in Ce-terminated surface. Some features originate from the Ce$^{(\text{B})}$ electron states can be observed in high energy range.
It also suggests that for the simulation of ARPES, the surface effect should be taken into consideration properly in other materials like CeRh$_2$Si$_2$.

\section{Conclusion}
In summary, we investigate the electronic structures of Si- and Ce-terminated surface of \CRS using first-principles approaches.
We have revealed three key aspects of surface effect on the electronic structures of CeRh$_2$Si$_2$.
Firstly, the relaxation of surface structure changes the dispersion of band structure, thus it adjusts the relative position of some high symmetry points with Fermi level. The enhancement of Kondo resonance is also be observed especially of $f_{7/2}$. From the changing of electron density after the surface relaxation, we believe that it is the enhancement of the bonding between layers renormalized the bands. More precise electric structure can be obtained with the relaxation of surface structure.
Secondly, the hybridization between $4f$ and $c$-electrons decreases from bulk to surfaces Ce atoms obviously, which suppresses the strength of Kondo peaks and shifts the spin-orbit peak position of surface $4f$ electrons.
Thirdly, the crystal electric field of outmost Ce atoms is different from the bulk, especially for Ce$^{(\text{S})}$ in Ce-terminated case whose $f_{5/2}$ orbitals are nearly degenerate like an isolate atom. By considering the crystal electric field effect on Ce $4f$ electrons, we have well reproduced the experimental ARPES results. All these simulation results strongly suggest that the surface has vital influence on the \CRS electronic properties, and a fully self-consist structure and electronic simulation on the DFT+DMFT may be needed for the investigation on other strongly correlated materials.

It should also be emphasized that the comparison of experimental and simulation DOS shows that the most ARPES information can be attributed to the surface electrons, but the information of bulk electrons also appear where the strength of surface DOS is weak. This may provide a different point of view to the interpretation of ARPES results.

\section{Acknowledgement}
We thank Guang-Ming Zhang, Jian-Zhou Zhao, Xing-Yu Gao, Huan Li, Dan Jian, Ming-Feng Tian, Yin Zhong and Fa-Wei Zheng for helpful discussions. The work was supported by the Science Challenge
Project (NO.TZ2018002 and NO.TZ2016001), National Nature Science Foundation of China (NO.U1930401,
NO.12004048 and NO.11974397), the National Key Research and Development Program of China (No.2017YFA0303104) and the Foundation of LCP. We thank the Tianhe platforms at the National Supercomputer Center
in Tianjin.

\section{Author Contributions}
H.-F. Song and Y. Liu conceived and supervised the project. Y.-C. Wang, Y.-J. Xu and Y. Liu performed the numerical simulations. All authors analysed and discussed the results. Y.-C. Wang, Y. Liu, Y.-J. Xu, X.-J. Han and H.-F. Song wrote the manuscript, with contributions from all the authors.

\clearpage

\end{document}


\title{Supplementary Materials to ``First-principles study on the role of surface in the heavy fermion compound CeRh$_2$Si$_2$''}

\author{Yue-Chao Wang}
\affiliation{Laboratory of Computational Physics, Institute of Applied Physics and Computational Mathematics, Beijing 100088, China}

\author{Yuan-Ji Xu}
\affiliation{Beijing National Laboratory for Condensed Matter Physics, Institute of Physics, Chinese Academy of Science, Beijing 100190, China}

\author{Yu Liu}
\email{liu\_yu@iapcm.ac.cn}
\affiliation{Laboratory of Computational Physics, Institute of Applied Physics and Computational Mathematics, Beijing 100088, China}

\author{Xing-Jie Han}
\affiliation{Beijing National Laboratory for Condensed Matter Physics, Institute of Physics, Chinese Academy of Science, Beijing 100190, China}

\author{Xie-Gang Zhu}
\affiliation{Science and Technology on Surface Physics and Chemistry Laboratory, Jiangyou, Sichuan 621908, China}

\author{Yi-feng Yang}
\affiliation{Beijing National Laboratory for Condensed Matter Physics, Institute of Physics, Chinese Academy of Science, Beijing 100190, China}
\affiliation{School of Physical Sciences, University of Chinese Academy of Sciences, Beijing 100190, China}
\affiliation{Songshan Lake Materials Laboratory, Dongguan, Guangdong 523808, China}

\author{Yan Bi}
\affiliation{Center for High Pressure Science and Technology Advanced Research, Beijing 100094, China}

\author{Hai-Feng Liu}
\affiliation{Laboratory of Computational Physics, Institute of Applied Physics and Computational Mathematics, Beijing 100088, China}

\author{Hai-Feng Song}
\email{song\_haifeng@iapcm.ac.cn}
\affiliation{Laboratory of Computational Physics, Institute of Applied Physics and Computational Mathematics, Beijing 100088, China}

\date{\today}
\maketitle

Our supplemental information is organized as follows. Sec.I shows the convergence tests results. Sec.II gives the detailed results of surface crystal structure relaxation and the discussion on the performance of difference methods. Sec.III shows our prediction results of band structure on other k-path. Sec.IV displays the band structure with crystal effect. Sec.V shows the variation of Kondo resonance peaks with surface relaxation.

\section{CONVERGENCE TESTS}

\begin{figure*}[h]
\centering
\includegraphics[width=0.8\textwidth]{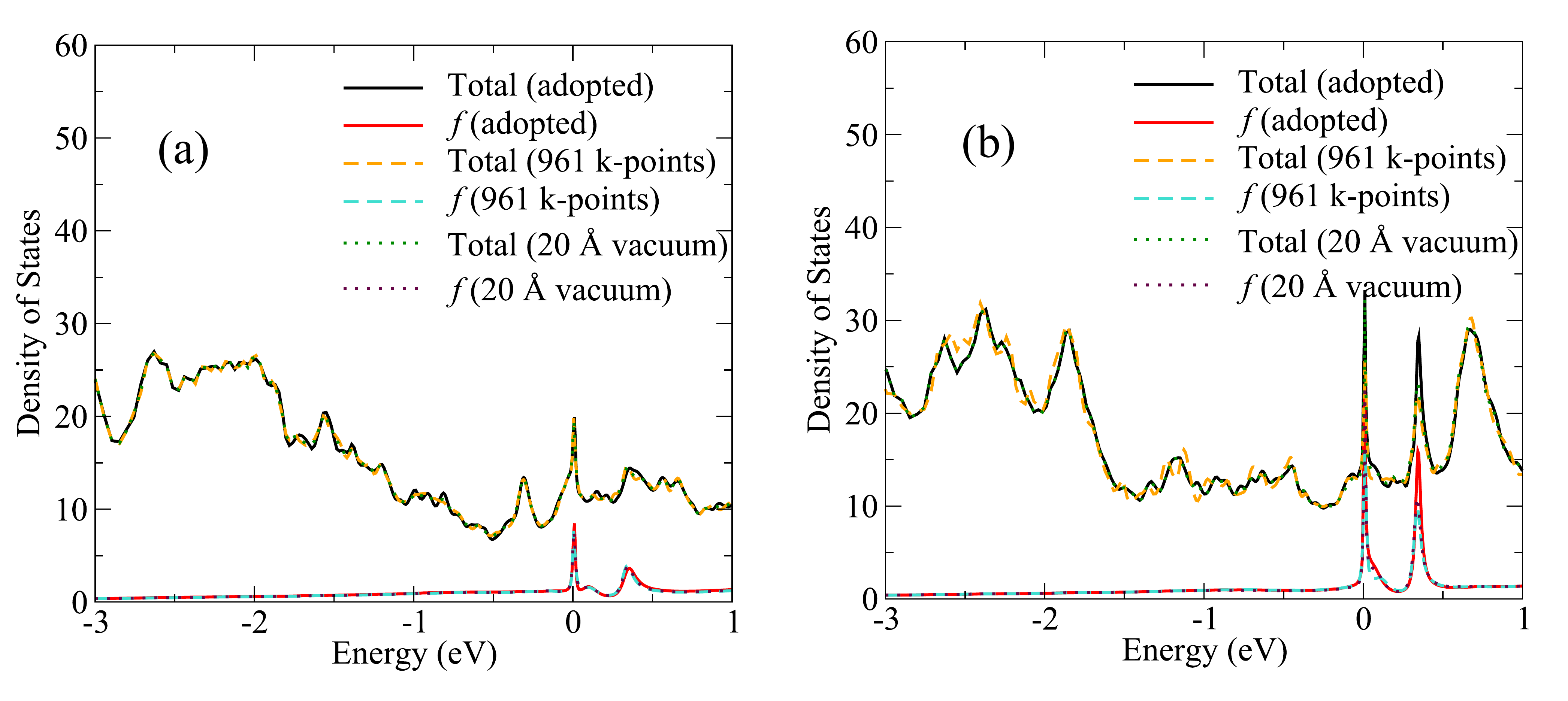}
\caption{\label{fig:conv}The convergence tests of k-points number and vacuum thickness in DFT simulation of slab models. Figure (a) and (b) are DOS (total DOS and PDOS of $f$-electrons) of convergence test in Si- and Ce-terminated cases respectively. The solid lines represent the parameters adopted in our work (625 k-points and 15 \AA { }vacuum), the dash line represent the simulation with denser k-mesh (961 k-points and 15 \AA { }vacuum), and dot line represent the simulation with thicker vacuum (625 k-points and 20 \AA { }vacuum).}

\end{figure*}
In our work, the $25\times25\times1$ mesh is used to sample the k-space of Si- and Ce-terminated slab, and a 15 \AA{ }thick vacuum is added to the slab model. To test the convergence of k-points number and vacuum thickness, we perform simulations with denser k-mesh and thicker vacuum, and the DOS (PDOS) results are shown in Fig.\ref{fig:conv}. The DOS results from denser k-mesh and thicker vacuum are almost the same, which indicates that the parameters used in this work are suitable for our purpose.



\section{COMPARISON BETWEEN DIFFERENT METHODS}

Table \ref{tab:STR} lists the variation of atom coordinate after the relaxation with DFT and DFT+$U$. The DFT and DFT+$U$ results are almost the same with each other in Si-terminated case. In Ce-terminated, the proportion of difference referring to the height of the sandwich shape sub-layer Rh-Si-Ce-Si-Rh is 0.3\% at surface Ce atom. It may be caused by the atom-like nature of surface Ce in Ce-terminated case, and this nature can be described more properly by DFT+$U$ than DFT. The results above confirm us that for CeRh$_2$Si$_2$ the on-site correlation correction may not much crucial for crystal structure relaxation. Considering the surface Ce in Ce-terminated case, and the crystal structure is relaxed at DFT+$U$ level.

\begin{figure*}[htp]
\centering
\includegraphics[width=0.8\textwidth]{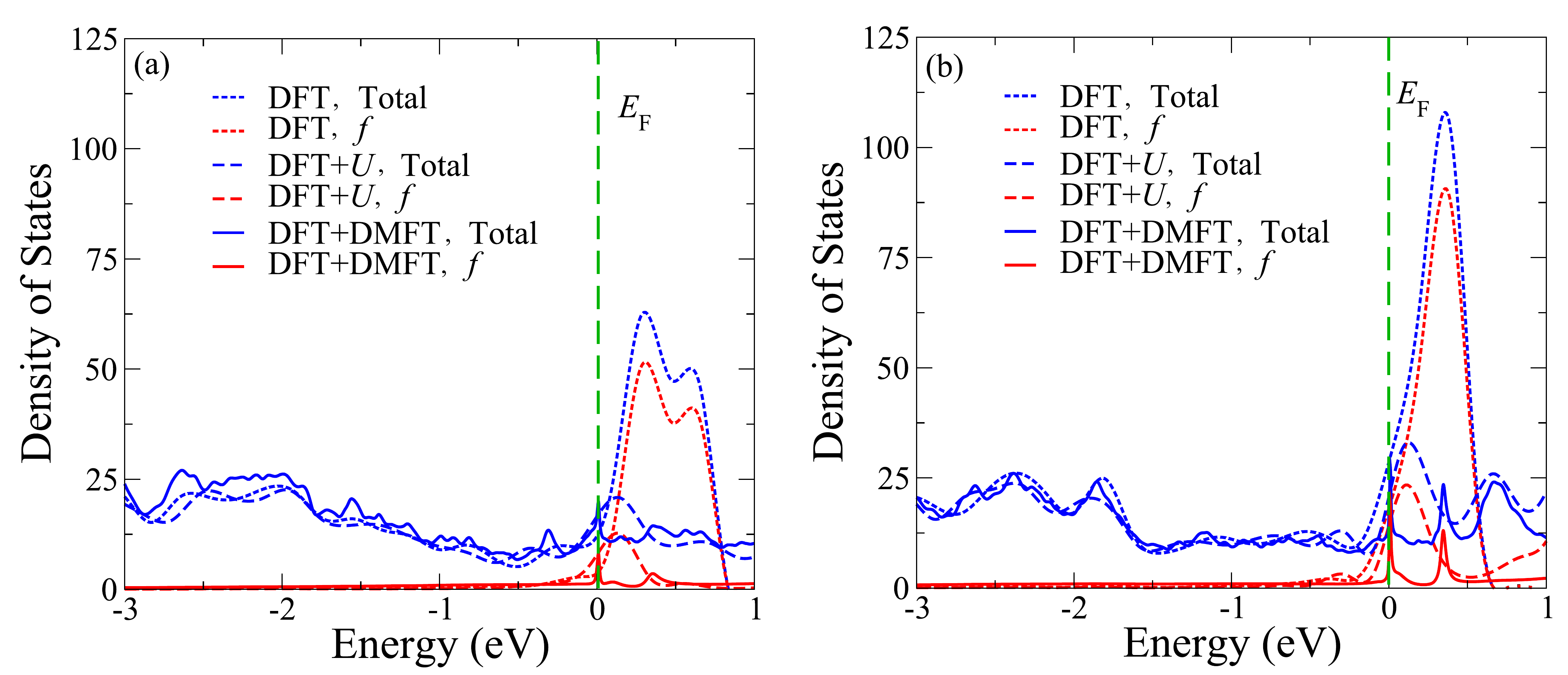}
\caption{\label{fig:dos}Density of states of $f$-electron (red) and total electron (black) from DFT (dot line), DFT+$U$ (dash line) and DFT+DMFT (solid line) in Si- (a) and Ce- (b) terminated cases.}

\end{figure*}

To investigate the similar crystal relaxation results of CeRh$_2$Si$_2$ in DFT and DFT+$U$, the density of states (DOS) from DFT, DFT+$U$ and DFT+DMFT of Si- and Ce-terminated cases are shown in Fig.\ref{fig:dos}. For particular this material, it is clearly shown that in the range far below Fermi level, the results of different methods are almost the same. The obvious differences appear at the area around Fermi level and it is mainly attributed to the $f$ electron of Ce. DOS of DFT+DMFT shows a sharp peak at Fermi level which is due to the $f_{5/2}$ resonance state, and another peak about 0.4 eV comes from the $f_{7/2}$ state. For the two $f$ peaks mentioned above are caused by dynamic correlated effect, no similar pattern is shown in the DOS of DFT or DFT+$U$.

In the results of DFT+$U$, a broaden $f$ peak can be also found around Fermi level, but this is different from the resonance state from DFT+DMFT calculation. The broaden peak of DFT+$U$ is caused by a half-filled $f$ orbital caused by the non-spin polarized condition used in +$U$ calculation. An orbital with 0.5 electron occupation in +$U$ approach will lead to nearly zero correction to the orbital, and the other $f$ orbitals without electron will be moved up by $U$/2 in energy to form the upper Hubbard band. Another evidence is that there is a very flat lower Hubbard band around -2.0 eV in DMFT simulation, while in DFT+$U$ simulation the corresponding $f$ states concentrate at the broaden peak at Fermi level. For DFT simulation results, almost all the $f$ states are concentrated in the range 0.0 eV to 1.0 eV, because of the lack of strongly correlated effect. As the differences of DOS from these three methods are not significant below Fermi level, it indicates that the prediction of total energy concerned properties (like crystal structure) may be not sensitive to the choice of method. \textcolor{black}{It should be mentioned that the opencore approach which means treating $f$ electrons as core states is also a conventional method to deal with lanthanides\cite{Poelchen2020}. However, as the interaction between $f$ and conduction electrons is of our concern in this work, we don't show the results from opencore simulation.}

\begin{table}[htp]
\centering
\caption{ The variation of atom position in Si- and Ce-terminated slabs calculated with DFT and DFT+$U$. The reference position is the corresponding atoms coordinates in bulk. The proportion of changing referring to the height of a sub-layer (Rh-Si-Ce-Si-Rh, 5.09 \AA) is shown in the parentheses.}
\label{tab:STR}
\begin{tabular*}{0.75\textwidth}{@{\extracolsep{\fill}} l|llrrr}
\hline\hline
Slab & \multicolumn{2}{c}{Si-terminated case} & \multicolumn{2}{c}{Ce-terminated case} &  \\
\hline
 Method &  DFT & DFT+$U$ & DFT & DFT+$U$ 	\\
 \hline
Layer-0 (Ce) (\AA)       & --&	--&	0.206 (4.0\%)&	0.186 (3.7\%)&  \\
Layer-1 (Si) (\AA)       & 0.202 (4.0\%)	&0.204 (4.0\%)	&-0.055 (-1.1\%)	&-0.042 (-0.8\%) &\\
Layer-2 (Rh) (\AA)	    &0.088 (1.7\%)	&0.087 (1.7\%)	&0.006 (0.1\%)&	0.006 (0.1\%) & \\
Layer-3 (Si) (\AA)	    & 0.060 (1.2\%)	&0.064 (1.2\%)	&0.094 (1.8\%)&	0.083 (1.6\%)  &\\
Layer-4 (Ce) (\AA)	       &0.066 (1.3\%)&	0.064 (1.2\%)&	0.000 (0.0\%)&	0.001 (0.0\%)&  \\
\hline\hline
\end{tabular*}
\begin{flushleft}
Note:

 $a.$ The value in the table is the difference of position between the relaxed atoms and unrelaxed atoms with the central layer fixed at the coordinate origin. The positive value means the atoms move to the center after the relaxation.\\

 $b.$ Layer-0 to Layer-4 represent the atomic layers from the outer to the inner. Layer-0 refers to the Ce terminated surface layer and there is no data for Si-terminated case. The element in every layer is given in the parentheses.
\end{flushleft}

\end{table}

\vbox{}
\vbox{}

\section{SPECTRAL FUNCTIONS ON ANOTHER  K-PATH}

\begin{figure*}[htp]
\centering
\includegraphics[width=0.8\textwidth]{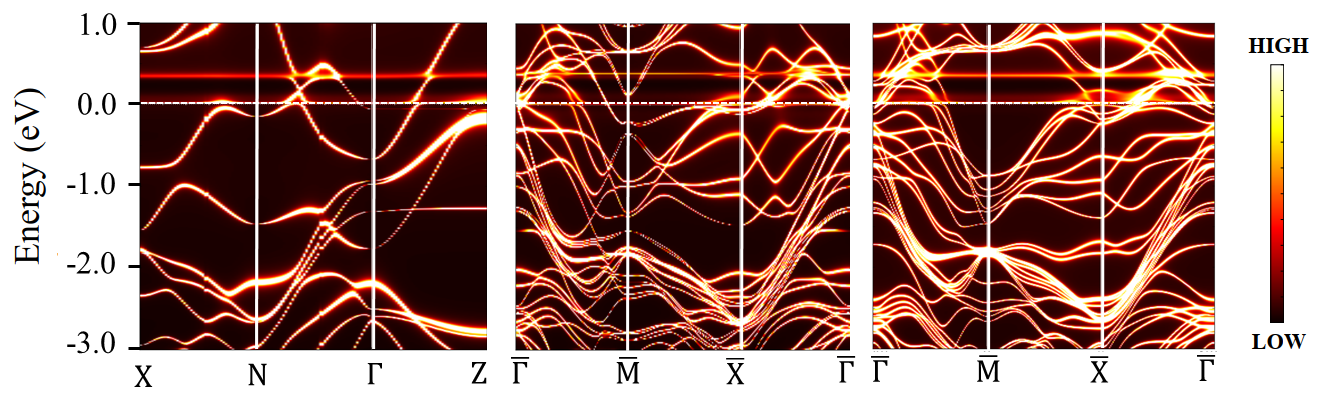}
\caption{\label{fig:pic1}The band spectral functions of bulk (left), Si-terminated case (middle) and Ce-terminated case (right) from DFT+DMFT calculations. The k-path here are along some common high symmetry points. Relaxed structures are used for Si- and Ce-terminated case.}

\end{figure*}

In Fig.\ref{fig:pic1} (left), the spectral function of bulk on a conventional k-path is shown. It is used to confirm our DFT+DMFT simulation is consistent with other works on CeRh$_2$Si$_2$ bulk\cite{Lu2018}. The Si- and Ce-terminated cases' spectral functions on another k-path are shown in Fig. (middle) and Fig. (right) for reference.

\begin{figure*}[htp]
\centering
\includegraphics[width=0.8\textwidth]{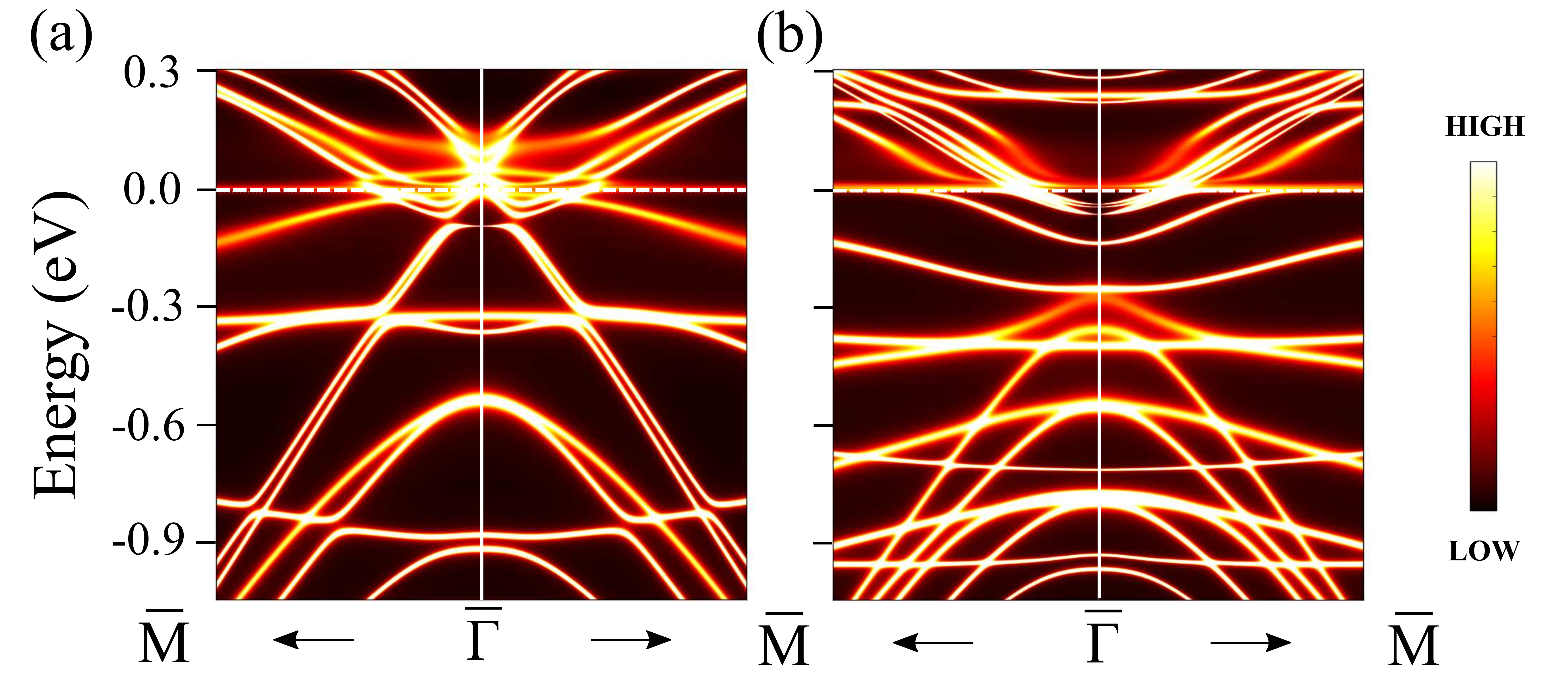}
\caption{\label{fig:zoomin}The zoom-in band spectral functions in Si-terminated case (a) and Ce-terminated case (b) of Fig.2(a) and Fig.2(c) respectively. The energy range are from -1.0 to 0.3 eV, and the k-path shows the first 20\% part along $\bar{\Gamma}$ to $\bar{\text{M}}$.}

\end{figure*}

\vbox{}
\vbox{}
\section{SPECTRAL FUNCTIONS WITH CRYSTAL EFFECT}

\begin{figure*}[htp]
\centering
\includegraphics[width=0.8\textwidth]{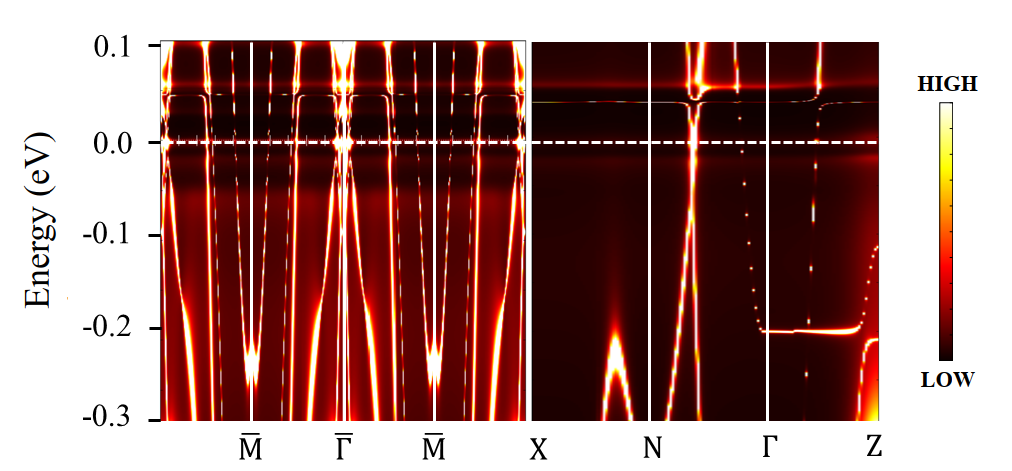}
\caption{\label{fig:pic2}The band spectral functions of Si-terminated case (left) and bulk (right) from DFT+DMFT calculations with crystal field effect.}

\end{figure*}

Fig.\ref{fig:pic2} displays the spectral function at Fermi level in Si-terminated case the bulk to show the effect of crystal effect. The spectral function results of Si-terminated is not so clear as the DOS in main text. It is because the electron states of outmost Ce and central Ce is mixed, but some characters can still be observed like the splitting of resonance peak at around -0.05 eV. For comparison a spectral function from bulk calculation is also shown in Fig.\ref{fig:pic2}. The characteristic f-state resonance bands in bulk also appear in the slab calculation, and it indicates that a simple simulation of spectral function simulation with slab or bulk may not reflect the experimental results properly, for the weight of bulk and surface states should not be the same.
\clearpage

\section{HYBRIDIZATION VARIATION WITH RELAXATION}

\begin{figure*}[htp]
\centering
\includegraphics[width=0.75\textwidth]{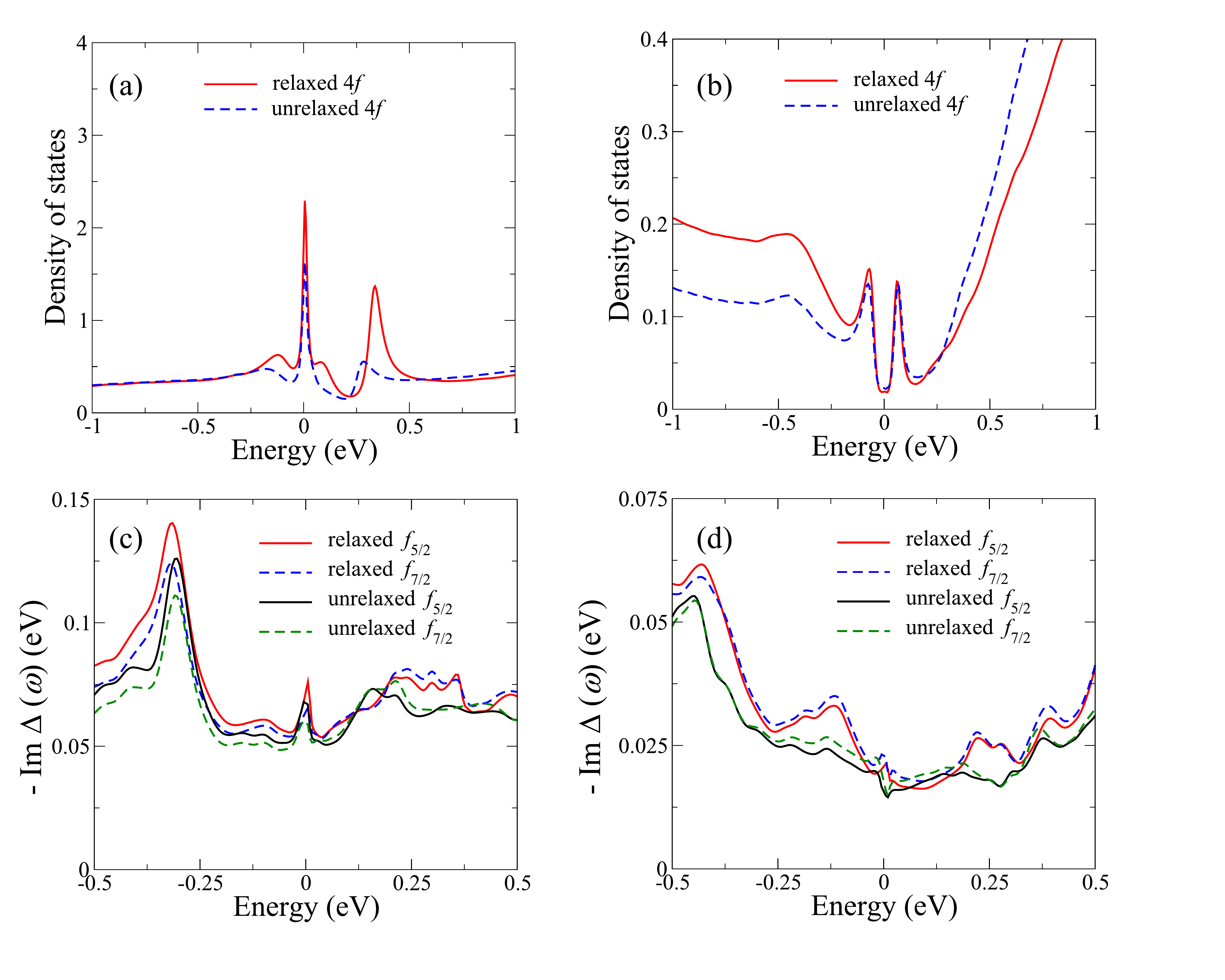}
\caption{\label{fig:hybrex}Density of states and imaginary part of hybridization functions in Si- and Ce-terminated case with and without relaxation. (a) and (b) are the DOS of surface Ce 4f-electrons in Si- and Ce-terminated cases respectively. The red solid lines are 4$f$-electron DOS with surface relaxation. The blue dash lines are 4$f$-electron DOS without surface relaxation. (c) and (d) show the imaginary part of hybridization function of Si- and Ce- terminated case respectively on real frequency. The solid red and black lines are for $f_{5/2}$ of surface Ce with and without relaxation. The dash blue and green lines are for $f_{7/2}$ of surface Ce with and without relaxation.}

\end{figure*}

Fig.\ref{fig:hybrex} shows the DOS of surface Ce 4$f$-electron with and without relaxation, and the corresponding imaginary part of hybridization functions on real frequency are also displayed. From the DOS results, the obvious enhancement of $f_{7/2}$ and the spin-orbit peaks can be observed. A slightly enhancement of Kondo peaks of $f_{5/2}$ also appears. The hybridization function results shows a weak enlargement of hybridization with the relaxation. It can be attributed to the contraction of surface atoms into the bulk, which strengthen the bonding with surrounding conduction electrons. However, the increasing of hybridization strength with the surface crystal structure relaxation is only a side effect caused by the cleavage of the surface, and it is much weaker than the variation of hybridization strength caused by surface itself as is shown in the text part III.B.

\vbox{}
\vbox{}
\clearpage

\vbox{}
\bibliographystyle{unsrt}